\documentclass{article}
\usepackage[T1]{fontenc} %
\usepackage[utf8]{inputenc} %
\usepackage{INTERSPEECH2020,amsmath,cite,url}
\usepackage{graphicx}
\usepackage{color}
\usepackage{booktabs}       %
\usepackage{multirow,array}
\usepackage{bbm}

\title{Unsupervised Cross-Domain Singing Voice Conversion}

\name{Adam Polyak$^{1,2}$\sthanks{ \hspace{0.1cm} The contribution of Adam Polyak is part of a Ph.D. thesis
research conducted at Tel Aviv University.}, Lior Wolf$^{1,2}$, Yossi Adi$^{1}$, Yaniv Taigman$^{1}$}
\address{
  $^1$Facebook AI Research \quad\quad\quad  $^2$Tel Aviv University}
\email{adampolyak@fb.com}

\begin{document}

\maketitle
\begin{abstract}
We present a wav-to-wav generative model for the task of singing voice conversion from any identity. Our method utilizes both an acoustic model, trained for the task of automatic speech recognition, together with melody extracted features to drive a waveform-based generator. The proposed generative architecture is invariant to the speaker's identity and can be trained to generate target singers from unlabeled training data, using either speech or singing sources. The model is optimized in an end-to-end fashion without any manual supervision, such as lyrics, musical notes or parallel samples. The proposed approach is fully-convolutional and can generate audio in real-time. Experiments show that our method significantly outperforms the baseline methods while generating convincingly better audio samples than alternative attempts.

\end{abstract}

\vspace{-0.1cm}
\section{Introduction}
\vspace{-0.1cm}
We tackle the audio synthesis task of singing voice conversion. In this task, a given template song is reproduced by another singer's voice. The conversion retains the content and musical expression of the template song. Singing voice conversion can aid in improving the vocal qualities of a given singing segment, create mimicry effects and even enable a single amateur singer to record an entire chorus in their home studio. 

Our method is inspired by recent work on speech voice conversion~\cite{polyak2019tts} and music synthesis~\cite{musicwavernn}. We combine melody extracted features with acoustic speech features and a powerful neural audio generation framework~\cite{parallelwavegan} to create a singing voice synthesizer. All features are extracted directly from the raw audio, therefore, our method does not require supervision in the form of a labelled dataset with lyrics and notes or a parallel dataset with singers singing the same songs. 

From a technical perspective, we present multiple contributions: (i) introducing an audio generation framework which employs task-related perceptual losses to improve the quality of the generated audio, (ii) the first singing voice conversion method, as far as we know, conditioned on sine-excitation of the song's melody, together with intermediate features of a speech recognition network, and (iii) presenting a speaker invariant singing voice conversion method trained on voices from either speech or singing datasets (i.e., mimicking in singing either speaking or singing voices), with either single or multiple identities.

\vspace{-0.1cm}
\section{Related work}
\vspace{-0.1cm}
\noindent{\bf Neural Audio Generation\quad} Recent advancements in neural audio generation enabled computers to generate natural sounding speech and music. Autoregressive models, such as WaveNet~\cite{wavenet} and SampleRNN~\cite{samplernn} generate high-quality audio in the waveform domain, one sample at a time, resulting in slow inference. WaveRNN~\cite{wavernn} enabled fast synthesis by training a compact neural network further optimized via sparsification. Feed-forward networks were suggested to further speed up inference, via knowledge distillation from an autoregressive teacher-model~\cite{parallelwavenet,ping2018clarinet, kim19flowavenet}. WaveGlow~\cite{waveglow} trained a feed-forward network without knowledge distillation. Recently, generative adversarial networks (GANs)~\cite{melgan, parallelwavegan} were able to match the quality of autoregressive and large feed-forward models. ParallelWaveGAN~\cite{parallelwavegan} trained a mel-inversion network, by combining adversarial networks with a multi-scale spectral loss. Other methods, employed multi-scale spectral loss to train convolutional neural networks for spectrogram inversion~\cite{arik2018fast}, to directly predict the parameters of a differentiable synthesizer~\cite{engel2020ddsp} and for speech synthesis with sine excitation as input~\cite{wang2019neural}.

\noindent{\bf Singing Synthesis and Conversion\quad} Previous singing synthesis methods applied unit selection methods~\cite{bonada2016expressive} or HMM based parametric methods~\cite{saino2006hmm, oura2010recent, nakamura2014hmm}. Neural networks were used to synthesize singing~\cite{singing}, by training a WaveNet-like network conditioned on notes and lyrics to generate vocoder features. This method was later extended to synthesize new singers based on a few minutes of them singing~\cite{singing2}. Mellotron~\cite{valle2019mellotron}, presented a sequence to sequence architecture conditioned on text and pitch to synthesize singing, without training on a singing dataset. Recently, non-autoregressive models were used for singing synthesis. Feed-forward transformers~\cite{singing3} removed the constraint of time-aligned phonemes, by employing a duration model. Generative adversarial networks~\cite{chandna2019wgansing}, conditioned on pitch and time-aligned phonemes, predicted a singing spectrogram in a single pass instead of a frame-by-frame prediction.

Initial methods for the task of singing voice conversion~\cite{kobayashi2015statistical,kobayashi2014statistical,conf/interspeech/VillavicencioB10}, relied on parallel datasets composed of paired samples of singers singing the same piece. The method of Unsupervised Singing Voice Conversion~\cite{usvc} learned to convert between a fixed set of singers without relying on a parallel-dataset. This was done by learning singer-agnostic features via a domain confusion term on the singer identity. PitchNet~\cite{pitchnet} further improved the method, by applying an additional domain confusion term on the pitch, but was only applied on conversion between a fixed group of male singers. 

Very recently, variational autoencoders were used to convert between singers and their vocal techniques~\cite{VAEsinging}, learned from non-parallel corpora. However, their method was limited to mel-spectrograms, thereby bounding the audio quality. A method employing an automatic speech recognition (ASR) engine~\cite{snap} was used for singing voice conversion. The ASR extracts phonemes probabilities, which the model then converts to the acoustic features of the target singer. The method was demonstrated on a many-to-one scenario and was only trained on a singing dataset, while our method is applicable both to many-to-many conversion and can also be trained on either speech or singing domains.  Similar to us, ~\cite{singwavernn} used WaveRNN conditioned on phoneme probabilities, pitch and a speaker i-vector to generate singing audio in the waveform-domain. Our method differs by (i) using a non-autoregressive model for real-time generation, and (ii) the usage of perceptual losses which, as we demonstrate, greatly boost the performance of our method.

\begin{figure*}[t!]
  \centering
 \begin{tabular}{c|c}
  \includegraphics[width=0.67\textwidth,trim={10 40 10 50},clip]{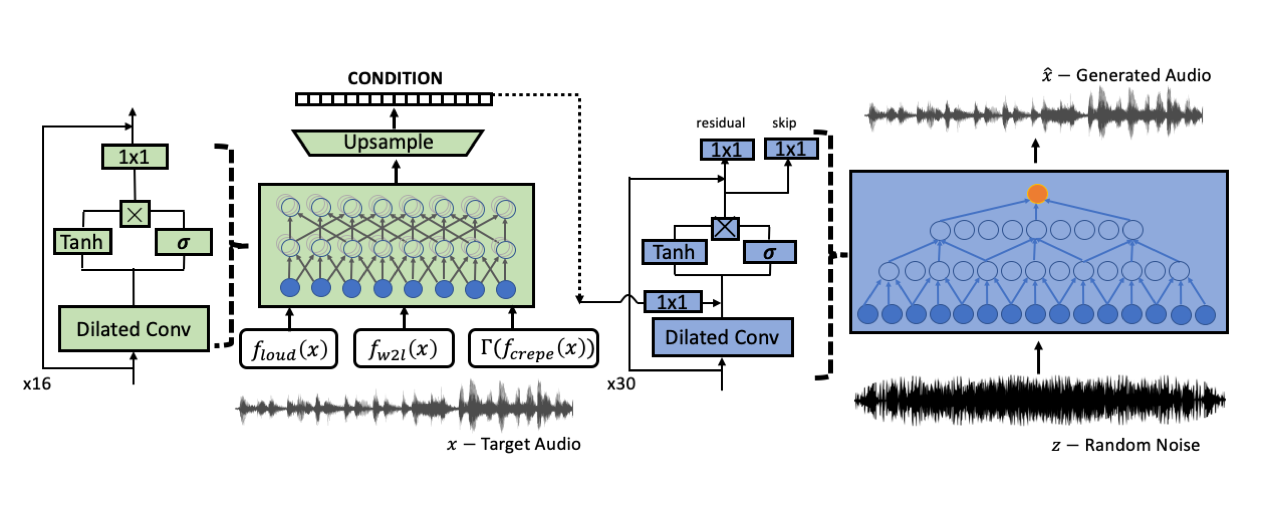} &
  \includegraphics[width=0.28\textwidth,trim={0 0 0 10},clip]{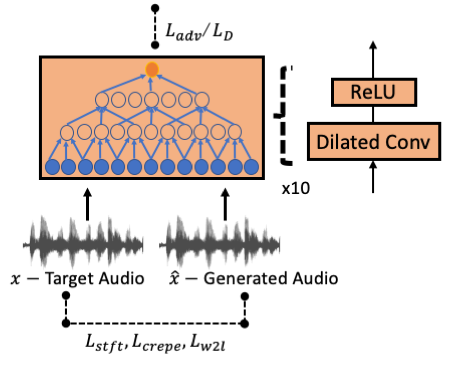} \\
 (a) & {\color{black}(b)} \\
 \end{tabular}
\caption{\textbf{Proposed GAN architecture.} (a) Generator architecture. Musical and speech features are extracted from a singing waveform $(f_{loud}(x), f_{w2l}(x), \Gamma(f_{crepe}(x)))$ and passed through context stacks (colored green). The features are then concatenated and temporally upsampled to match the audio frequency. The joint embedding is used to condition a non-causal WaveNet (colored blue), which receives random noise as input. (b) Discriminator architecture. Losses are drawn with dashed lines, input/output with solid lines. The discriminator (colored orange) differentiates between synthesized and real singing. Multi-scale spectral loss and perceptual losses are computed between matching real and generated samples.}
\label{fig:arch}
\vspace{-0.4cm}
\end{figure*}

\vspace{-0.1cm}
\section{Method}
\vspace{-0.1cm}
The proposed model is based on a Generative Adversarial Network, with a generator network $G$ and a discriminator network $D$. The model is conditioned on both speech and musical features, while in the multi-singer generation case, it is additionally conditioned on a learned singer identity vector.
Each feature set is forwarded via a separate context-stack, similar to~\cite{musicwavernn, midi2wave2midi}, before feeding it to $G$. The generator is a non-autoregressive  WaveNet, which generates the audio waveform directly from a random noise vector. 
Fig.~\ref{fig:arch} depicts the architecture.

\noindent{\bf Input Features\quad}
We denote the domain of audio samples by $\mathcal{X} \subset \mathbb{R}$. The representation for a raw speech signal is therefore a sequence of samples $\bm{x} = (x_1,\ldots, x_T)$, where  $x_t\in\mathcal{X}$ for all $1\leq t \leq T$. The length of the input signal varies for different inputs, thus the number of input samples in the sequence, $T$, is not fixed. 
Given a training set of $n$ examples, $S = \{\bm{x}_i\}_{i=1}^n$, we would like to extract representations which are both speaker invariant, to enable better singer conversion, and independent of manual annotations, to utilize unlabelled data.

Recent works demonstrated the need of both linguistic and musical features~\cite{singing, singing2, snap} in the context of singing generation. As a result, we extract both the loudness measure~\cite{loudness} and the fundamental frequency (F0) as the musical features. Loudness is represented by the log-scaled A-Weighting of the power spectrum, $f_{loud}(\bm{x})$, while F0 is extracted using CREPE~\cite{crepe}, denoted by $f_{crepe}(\bm{x})$, similarly to~\cite{musicwavernn}. We experimented with different representations of F0, such as: octave, note, etc., and achieved similar performance.

In preliminary experiments, we observed that using F0 as input produces inconsistent shakes in the pitch of the generated samples. Therefore, we turn into conditioning on a synthesized melody generated from the F0 instead. The melody is synthesized via a single sinusoid sine-excitation, denoted by $\Gamma(f_{crepe}(\bm{x}))$.

For speech features, we follow~\cite{polyak2019tts} and utilize an intermediate representation from a pre-trained acoustic model optimized for the task of Automatic Speech Recognition (ASR) as an additional input to the model. Specifically, we use the public implementation~\cite{jasper} of Wav2Letter~\cite{wav2letter}, denoted by $f_{w2l}(\bm{x})$. Since an ASR network is speaker-agnostic by design~\cite{adi2019reverse}, our method does not require any disentanglement terms or domain-specific (speaker) confusion terms.

Finally, we concatenate and upsample the features in the temporal domain to match the audio frequency.

\smallskip
\noindent{\bf Single Singer Objective Function\quad}
We follow the least-squares GAN\cite{mao2017least} setup where the discriminator and generator would like to minimize the following terms,
\begin{equation}
\label{eq:discriminator}
\begin{aligned}
    &L_D(D, G, S) = \sum_{\bm{x} \in S}{[|| 1 - D(\bm{x}) ||_2^2 + || D(\hat{\bm{x}}) ||_2^2]}\\
    &L_{adv}(D, G, S) = \sum_{\bm{x} \in S}|| 1 - D(\hat{\bm{x}}) ||_2^2
\end{aligned}
\end{equation}
accordingly. $S$ is the set of samples, $\hat{\bm{x}}= G(z, E(\bm{x}))$ is the audio sample synthesized from a random noise vector sampled from a uniform distribution $\bm{z}\sim U(0,1)$, and the concatenated features $E(\bm{x}) = [f_{loud}(\bm{x}), f_{w2l}(\bm{x}), \Gamma(f_{crepe}(\bm{x}))]$. 

In addition, we include a reconstruction loss to further improve optimization stability. We note that two audio samples might be perceptually similar while being the exact opposite in the waveform representation, e.g., in the case of a simple phase inversion (multiply by -1). To mitigate this, we include a spectral amplitude distance loss~\cite{oord2017parallel, arik2018fast} in multiple FFT resolutions~\cite{wang2019neural, parallelwavegan, engel2020ddsp}. The spectral amplitude distance loss, for a given FFT size $m$, is defined as follows:
\begin{equation}
    L_{\text{recon}}^{(m)}(G, S) = \sum_{\bm{x} \in S}{ \left[ \frac{\| \mathcal{S} - \hat{\mathcal{S}} \|_F}{\| \mathcal{S} \|_F} + \frac{\| \log\mathcal{S} - \log\hat{\mathcal{S}} \|_1}{N} \right]}
\end{equation}
where $\| \cdot \|_F$ and $\| \cdot \|_1$ denotes the Forbenius and the $L_1$ norms, $\mathcal{S}=|\text{STFT}(x)|$ and $\hat{\mathcal{S}}=|\text{STFT}(\hat{x})|$ denotes the Short-time Fourier transform magnitudes of the original and synthesized samples respectively, and $N$  the number of elements. The first term of the expression emphasizes spectral peaks, while the second penalizes silent sections of the audio. The multi-resolution loss is defined as the sum of the above loss for multiple scales:
\begin{equation}
    L_{recon}(G, S) =  \frac{1}{|M|}\sum_{m \in M} L_{recon}^{(m)}(G, S)
\end{equation}
where $M = [2048, 1024, 512, 256, 128, 64]$.

Lastly, to further improve the generation quality, we add perceptual losses~\cite{Johnson2016Perceptual} on top of the generator output. Specifically, we compute the $l_1$-distance between intermediate activations of the ASR network, $h_{w2l}$ and the CREPE network, $h_{crepe}$ as follows:
\begin{equation}
\begin{aligned}
&L_{crepe}(G, S) = \sum_{\bm{x} \in S} \| h_{crepe}(\bm{x}) - h_{crepe}(\hat{\bm{x}}) \|_1\\
&L_{w2l}(G, S) = \sum_{\bm{x} \in S} \| h_{w2l}(\bm{x}) - h_{w2l}(\hat{\bm{x}}) \|_1
\end{aligned}
\end{equation}
Overall, the optimization loss for the generator, $G$, is defined as: 
\begin{equation}
\label{eq:generator}
\begin{aligned}
L_G(G, D, S) = &  L_{recon}(G, S) + \alpha L_{adv}(G, D, S)\\
            & + \beta L_{crepe}(G, S) + \gamma L_{w2l}(G, S)
\end{aligned}
\end{equation}
where $\alpha, \beta, \gamma$ are weight factors to balance the contribution of each loss term.

\noindent{\bf Multi-singer Training Losses\quad}
In the multi-singer regime, we include a speaker embedding $\bm{v}_i$ as an additional input to $G$. The speaker embeddings are learned during training and stored in a Look Up Table. Then, the reconstruction of a sample, $\bm{x}_i$, pronounced by speaker $i$ is updated to be $\bm{\hat{x}}_i^{i}= G(\bm{z}, E(\bm{x}_i), \bm{v}_i)$.

Moreover, we introduce two additional training schemes. The first one is performed by converting a sample from singer $i$ to singer $j$, while omitting the reconstruction loss. The second one introduces novel virtual training samples by creating parallel samples using back-translation~\cite{sennrich2015improving} and mixup~\cite{zhang2017mixup}. This was previously shown by~\cite{usvc} to improve singing voice conversion. 

These additional objective functions for unaligned samples are defined as follows, 
\begin{equation}
    \begin{aligned}
    L_{D}^{unaligned}(D, G, S)& = \sum_{\bm{x}_i \in S}{\left[\| 1 - D(\bm{x}_i) \|_2^2 + \| D(\hat{\bm{x}}_j^i) \|_2^2\right]}\\
    L_{G}^{unaligned}(G, D, S)& = \alpha L_{adv}(G, D, S) + \beta L_{crepe}(G, S) \\
                      & + \gamma L_{w2l}(G, S)
\end{aligned}
\end{equation}
where $\hat{\bm{x}}_j^i=G(\bm{z}, E(\bm{x}_i), \bm{v}_j)$. 
Note the reconstruction loss is omitted, since we do not have the target sample of singer $j$ singing sample $\bm{x}_i \in S$.  

To virtually simulate unseen singers we follow the mixup scheme while generating a conversion to a new virtual singer. Specifically, we use a convex combination of two different singers embeddings, $\bm{v}_j$ and $\bm{v}_j'$, as:
\begin{equation}
    \bm{u} = \nu \bm{v}_j + (1 - \nu) \bm{v}_{j'}
\end{equation}
where $\nu \sim U[0, 1]$ is drawn from the uniform distribution. Sample $\bm{x}_{\bm{u}}^j = G(\bm{z}, E(\bm{x}_j), \bm{u})$ is then generated from $\bm{x}_j$, and translated back to singer $j$ as follows: $\bm{\hat{x}}_{\bm{u}}^j = G(\bm{z}, E(\bm{x}_{\bm{u}}^j), \bm{v}_j)$.
The produced set of artificial examples is denoted as $S_{mixup}=\{ (\bm{x}_j, \bm{\hat{x}_u^j})\}$. %
Finally, the discriminator and generator are optimized by minimizing the loss over supervised, unaligned and virtual samples: 
\begin{equation}
\begin{aligned}
L_D^{multi}(G, D) = & L_D(G, D, S) + L_D^{unaligned}(G, D, S)\\ 
                    & + L_D(G, D, S_{mixup}) \\
L_G^{multi}(G, D) = & L_G(G, D, S) + L_G^{unaligned}(G, D, S) \\ 
                    & + L_G(G, D, S_{mixup})
\end{aligned}
\end{equation}
Note that in the mixup setting, the $\bm{\hat{x}} = \bm{\hat{x}}_{\bm{u}}^j$. 

\noindent{\bf Architecture\quad}
Generator $G$, is based on a non-causal WaveNet architecture~\cite{denoisingwavenet, parallelwavegan}. It receives as input a random noise vector sampled from a uniform distribution, $\bm{z} \sim U(0,1)$ and the input features described above. Each input feature ($f_{loud}, f_{w2l} \text{ and } f_{crepe}$) is passed through a separate convolutional stack~\cite{midi2wave2midi, musicwavernn}, which is composed of two blocks of eight non-causal convolutional layers. The layers in each block have an exponentially increasing dilation rate. Each layer employs 128 filters and a kernel-size of 3. The features are then upsampled by a series of interleaved nearest neighbor upsampling and convolutional layers. Once temporally-aligned to the audio, the features are concatenated to form the conditioning signal.

The generator is composed of a series of 30 non-causal layers ordered in three blocks. The dilation rate in a single block of layers is exponentially increasing. Thus, the model has a receptive field of 3,072 samples, which means each sample is generated based on a window of $96 ms$ in future and past directions. Each layer has 128 residual-channels, 128 skip-channels and a kernel-size of 3. Our model achieves inference speed of 10.14 times faster than real-time, on a single Tesla V100 GPU.

The discriminator is composed of ten layers of 1-D convolutions followed by a leaky-ReLU activation with a leakiness of 0.2, with a linearly increasing dilation rate. Each convolution layer consists of 128 filters with a kernel-size of 3. The discriminator outputs a prediction for each time-step in the input audio signal. The loss is computed by averaging across the time-domain. Both the discriminator and generator apply weight normalization~\cite{weightnorm} on all layers.

\begin{table*}[t!]
\centering
\caption{Test scores for singing voice conversion. For MOS, SIM, identification, higher is better. For VDE and FFE -- lower.}
\vspace{-.1cm}
\label{tab:results}
\begin{tabular}{@{}llc@{~}c@{~}cc@{~}c@{~}c@{}}
\toprule
Dataset & Method & MOS~$\uparrow$ & SIM~$\uparrow$ & Identification~$\uparrow$  & VDE~$\downarrow$ & FFE~$\downarrow$\\
\midrule
Source Singing & Ground Truth & 4.10$\pm$0.84 & 86.11\% & 100\% & --- & --- \\
\midrule
\multirow{3}{*}{LJ} 
& Mellotron &  3.79$\pm$1.06 & 59.04\% & 76.47\% & 8.35\% & 9.50\% \\
& Ours  &  4.06$\pm$0.81 & 60.46\% & 97.87\%  & 4.19\% & 5.51\% \\
& Ground Truth  &  4.51$\pm$0.70 & 70.83\% & 97.91\% & --- & --- \\ 
\midrule 
\multirow{3}{*}{VCTK} 
& Mellotron     & 3.14$\pm$ 0.82 & 70.00\% & 57.67\% &  15.02\% & 16.62\% \\
& Ours      & 3.88$\pm$ 0.56 & 78.57\% & 96.41\% & 6.98\% & 7.72\% \\
& Ground Truth  & 4.35$\pm$ 0.74 & 75.00\% & 99.28\% & --- & --- \\
\midrule
\multirow{3}{*}{LCSING} 
& USVC  &  3.52$\pm$0.91 & 38.24\% & 8.51\% & 8.19\% & 11.46\% \\
& Ours              &  3.81$\pm$0.92 & 67.74\% & 100\%  & 4.00\% & 4.80\% \\
& Ground Truth    &  3.95$\pm$0.73 & 77.78\% & 100\%  & --- & --- \\
\midrule
\multirow{3}{*}{NUS-48E} 
& Mellotron  & 3.55$\pm$0.87 & 69.44\% & 60.11\%  &  6.59\% & 8.02\% \\
& WGANSing  &  3.60$\pm$0.94 & 80.56\% & 92.27\%  &  3.85\% & 5.06\% \\
& USVC      &  3.78$\pm$0.85 & 61.90\% & 93.45\%  &  4.82\% & 20.80\% \\
& Ours      &  4.04$\pm$0.68 & 81.82\% & 97.02\%  &  2.47\% & 3.50\% \\
\bottomrule
\end{tabular}
\vspace{-.4cm}
\end{table*}

\vspace{-0.1cm}
\section{Experiments}
\label{sec:experiments}
\vspace{-0.1cm}
We perform a series of experiments to evaluate the proposed method against several baselines. We experimented with generating singing using speech-only, singing-only and mixed datasets. We explore both many-to-one conversion on a large single identity corpus and many-to-many conversion using learned identities from a corpus with a variable amount of audio per identity. Moreover, we perform an extensive ablation study to better understand the contribution of each component. Audio samples are available online at \url{https://singing-conversion.github.io/}, as well as in the supplementary material.

\noindent{\bf Datasets\quad} We report results on several datasets. {\em LJ}~\cite{ljspeech} is a large single speaker speech corpus with approximately 24 hours of audio recording. {\em  LCSING} is a studio recordings of a single professional singer~\cite{lucilecrew}, which was filtered using an off-the-shelf voice activity detector and contains 3 hours and 40 minutes of very expressive and high dynamic range recordings, including some melodic singing without lyrics. For the multi-speaker experiments, we use the speech corpus, {\em VCTK}~\cite{vctk}, to learn 109 singers with 44 hours of audio recordings. Finally, the {\em NUS-48E}~\cite{duan2013nus} dataset, which includes six male singers and six female singers, has both singing and reading of four songs per voice, resulting in a total of ~15 minutes per singer. All audio was down-sampled to 16kHz with a single channel. All datasets were randomly split according to a 80\%/10\%/10\% of train/val/test partitions.

\noindent{\bf Hyperparameters\quad} We train our models for 800K steps, with a batch-size of 8 one second long audio segments. We use RADAM~\cite{radam} optimizer with a learning rate of 0.0001. The learning rate is halved every 200K steps. The discriminator joins the training process after 100K steps and the perceptual losses after 50K steps. For the CREPE perceptual loss, we use the intermediate-activation before the final sigmoid activation. For the ASR network loss, we use the output of the tenth convolutional block. We use $\alpha=4,\beta=1,\gamma=10$ for the weight factors in Eq.~\ref{eq:generator}. The model is trained with a mixup batch every 3 steps after 100K steps of training had passed.

\smallskip
\smallskip
\noindent{\bf Singing conversion\quad} All experiments were performed by converting singing recordings of the NUS-48E to the target identities learned from the datasets described above. Therefore, experiments conducted on the LJ, LCSING and VCTK evaluate the methods' invariance to the \textit{input} voice identity, while experiments conducted on the NUS-48E dataset, evaluate the methods' performance while converting across a fixed set of singers.

Evaluation metrics are based on subjective and objective success metrics: (i) Mean Opinion Scores (MOS), human raters rate the naturalness of the audio samples on a scale of 1--5. Each experiment, included 40 randomly selected samples rated by 20 raters. (ii) ABX testing for similarity, in which we present each rater with two audio samples A and B. The examples originate from two different singers. These two samples are followed by a third utterance X randomly selected to be from the same identity as A or B. Next, the rater must decide whether X has the same identity as A or B. We report the success rate across all raters. (iii) Automatic identification metric by training a multi-class classifier on the ground-truth training partitions of all datasets and reporting the success rate of the classifier, similar to~\cite{deepvoice2,taigman2017voice,nachmani2018fitting}. %
(iv) Voicing Decision Error (VDE)~\cite{nakatani2008method}, which measures the portion of frames with voicing decision error,
(v) F0 Frame Error (FFE)~\cite{chu2009reducing}, measures the percentage of frames that contain a deviation of more than 20\% in pitch value or have a voicing decision error.%

Mellotron~\cite{valle2019mellotron} showed good results on singing generation by training on speech datasets. Therefore, we use it as the baseline for the experiments on speech datasets. Given a template singing sample, Mellotron extracts the rhythm, the alignment between text and spectral features, which is then used to generate the conversion. For emotive samples, like in the NUS-48E dataset, the rhythm extraction sometime failed. Therefore, we used a forced-aligner~\cite{mcauliffe2017montreal} to create a synthetic alignment map and replace the rhythm extracted by Mellotron with the synthetic one before generating the conversion.

For a fair comparison we do not report Mellotron results on LCSING dataset due to instability in the training caused by the following reasons. In the LCSING dataset there are only 40 minutes of transcribed segments and many of the recordings contain non-lexical vocables.

Our experiments on NUS-48E dataset involved baselines which showed convincing results: WGANSing~\cite{chandna2019wgansing} and Unsupervised Singing Voice Conversion (USVC)~\cite{usvc}. Both methods require multi-singer voice dataset for training. Therefore, we do not apply them on LJ and VCTK. For the single singer dataset, we use the same architecture as USVC but without the confusion term.

Table~\ref{tab:results} presents the results for all of the above models. On speech datasets, LJ and VCTK, our models outperform Mellotron, despite the latter utilizing the underlying text as input. In addition to subjective quality, our method predicts both voicing decision and pitch accuracy better than Mellotron. Results on LCSING, show that our method is better than the baseline and is able to generate recognizable samples. On a multi-singer dataset, NUS-48E, our method generates subjectively higher quality samples, which are more identifiable than the baselines.

\begin{table}[t!]
\centering
\caption{Ablation study on the LJ dataset.}
\vspace{-.2cm}
\label{tab:ablation}
\begin{tabular}{@{}l@{~}c@{~}c@{~}c@{~}c@{~}@{}}
\toprule
Method & MOS~$\uparrow$ &  VDE~$\downarrow$  & FFE~$\downarrow$\\
\midrule
F0 condition            &  2.37$\pm$0.93 &  7.20\% & 8.81\% \\
Melody condition        &  3.26$\pm$1.04 &  6.33\% & 7.90\% \\
+ ASR perceptual loss   &  3.36$\pm$0.96 &  6.47\% & 8.03\% \\
+ CREPE perceptual loss &  4.06$\pm$0.82 &  4.19\% & 5.51\% \\
\bottomrule
\end{tabular}
\end{table}

\noindent{\bf Ablation\quad}
We perform ablation for the suggested single singer training framework. Table~\ref{tab:ablation} summarizes the results. The comparison between the F0 condition and the Melody condition, shows that providing melody as input to the model reduces the overall error with regard to pitch generation. The addition of the ASR perceptual loss slightly improves the model MOS scores at the cost of slightly reducing the pitch metrics. Adding the CREPE perceptual loss adds a significant gain to the model performance across all metrics.

\vspace{-0.1cm}
\section{Conclusion}
\vspace{-0.1cm}
We present an unsupervised method that can convert a singing voice to a voice that is sampled either as speaking or singing. The method employs multiple pre-trained encoders and perceptual losses and achieves state of the art results on both objective and subjective measures. Conditioning the generator on a sine-excitation was shown to be beneficial while further improving the results. As future work, we would like to focus on temporal modification of the input singing to further match the style of the target singer.

\clearpage

\bibliographystyle{IEEEtran}
\bibliography{voice}

\end{document}